  \documentclass[twocolumn,PRL,aps,psfig,preprintnumbers,superscriptaddress]{revtex4}
\usepackage{graphicx}
\usepackage{epstopdf}
\usepackage{float}
\usepackage{color}
\usepackage{amsmath}

\begin{document}

\title{Magnetic properties of the itinerant ferromagnet LaCrGe$_{3}$ under pressure studied by $^{139}$La NMR}
\author{K. Rana}
\affiliation{Ames Laboratory, U.S. DOE,  Iowa State University, Ames, Iowa 50011, USA}
\affiliation{Department of Physics and Astronomy, Iowa State University, Ames, Iowa 50011, USA}
\author{H. Kotegawa}
\affiliation{Department of Physics, Graduate School of Science, Kobe University, Kobe 657-8501, Japan}
\author{R. R. Ullah}
\affiliation{Department of Physics, University of California, Davis, CA 95616, USA}
\author{E. Gati$\footnote[1]{Present address: Max Planck Institute for Chemical Physics of Solids, 01187 Dresden, Germany}$}
\affiliation{Ames Laboratory, U.S. DOE,  Iowa State University, Ames, Iowa 50011, USA}
\affiliation{Department of Physics and Astronomy, Iowa State University, Ames, Iowa 50011, USA}
\author{S. L. Bud'ko}
\affiliation{Ames Laboratory, U.S. DOE,  Iowa State University, Ames, Iowa 50011, USA}
\affiliation{Department of Physics and Astronomy, Iowa State University, Ames, Iowa 50011, USA}
\author{P. C. Canfield}
\affiliation{Ames Laboratory, U.S. DOE,  Iowa State University, Ames, Iowa 50011, USA}
\affiliation{Department of Physics and Astronomy, Iowa State University, Ames, Iowa 50011, USA}
\author{H. Tou}
\affiliation{Department of Physics, Graduate School of Science, Kobe University, Kobe 657-8501, Japan}
\author{V. Taufour}
\affiliation{Department of Physics, University of California, Davis, CA 95616, USA}
\author{Y. Furukawa}
\affiliation{Ames Laboratory, U.S. DOE,  Iowa State University, Ames, Iowa 50011, USA}
\affiliation{Department of Physics and Astronomy, Iowa State University, Ames, Iowa 50011, USA}

\date{\today}

 \begin{abstract} 

$^{139}$La nuclear magnetic resonance (NMR) measurements under pressure ($p = 0-2.64$ GPa) have been carried out  to investigate the static and dynamic magnetic properties of the itinerant ferromagnet LaCrGe$_3$. 
  $^{139}$La-NMR spectra for all measured pressures  in the ferromagnetically ordered state show a large shift due to the internal field induction $|$$B_{\rm int}$$|$ $\sim$ 4 T at the La site produced by Cr ordered moments.
  The change in $B_{\rm int}$ by less than 5\%  with $p$ up to 2.64~GPa indicates that the Cr 3$d$ moments are robust under pressure.
  The temperature dependence of NMR shift and $B_{\rm int}$ suggest that the ferromagnetic order develops below $\sim$ 50~K under higher pressures in a magnetic field of $\sim$ 7.2~T. 
  Based on the analysis of NMR data using the self-consistent-renormalization (SCR) theory, the spin fluctuations in the paramagnetic state well above $T_{\rm C}$ are revealed to be three dimensional ferromagnetic throughout the measured $p$ region. 

\end{abstract}

\maketitle

 \section{Introduction} 


 A quantum critical point (QCP), defined as a second order phase transition at absolute zero temperature $T$ = 0 K \cite{Sachdev2011} in itinerant ferromagnets has attracted much attention owing to the observation of a wide variety of intriguing phenomena such as unconventional superconductivity \cite{K5,K6,K7,Aoki2001,Huy2007}, non-Fermi liquid behavior \cite{Pfleiderer2001,Takashima2007,Maeda2018} and peculiar magnetic properties originating from quantum criticality \cite{Canfield2016,Brando2016} which can be controlled by tuning parameters such as pressure ($p$) and external magnetic field ($H$). 
   Interestingly,  it is now generally known that the ferromagnetic (FM) QCP in clean itinerant ferromagnets are avoided \cite{Brando2016}, although exceptions with  noncentrosymmetric metals having strong spin-orbit interaction are pointed out by the recent theoretical \cite{Kirkpatrick2020} and experimental \cite{Kotegawa2019, Shen2020} studies. 
 
 In most itinerant ferromagnetic systems,  when the second order paramagnetic (PM)-FM phase transition temperature ($T_{\rm C}$) is suppressed by the application of $p$, the order of the phase transition changes to the first order at the tricritical point (TCP)  before $T_{\rm C}$  reaches 0 K at the quantum phase transition (QPT), known as the avoided QCP \cite{Taufour2010,Kotegawa2011,Kabeya2012}. 
 When the PM-FM transition becomes of the first order at the TCP in the $p$-$T$ plane, 
the application of $H$ leads to a tricritical wing (TCW) structure in the $T$-$p$-$H$ three dimensional phase diagram  as found in UGe$_{\rm 2}$ \cite{Taufour2010,Kotegawa2011} and ZrZn$_{\rm 2}$\cite{Kabeya2012}. 
   A  PM-FM QCP  can also be avoided by the appearance  of an antiferromagnetic (AFM) ordered state under $p$ near the putative QCP, as actually observed in  CeRuPO\cite{Kotegawa2013,Lengyel2015},  MnP\cite{Cheng2015,Matsuda2016} and La$_5$Co$_2$Ge$_3$ \cite{Xiang2021}.
    In this case, no wing structure has been reported and the AFM state is suppressed by the application of moderate $H$.

  \begin{figure}[tb]
\includegraphics[width=6cm]{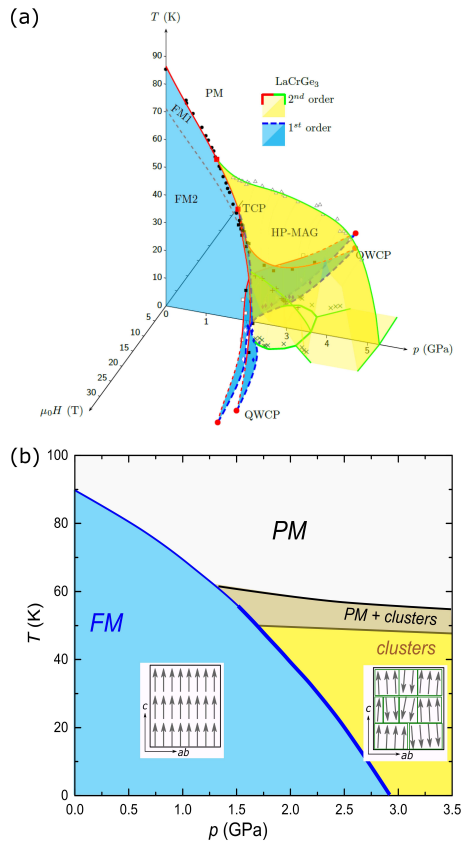} 
\caption{(a) Three dimensional temperature-pressure-magnetic field ($T$-$p$-$H$) phase diagram for LaCrGe$_3$ taken from Ref.~\cite{Kalu2017} where $H$ is applied parallel to the $c$ axis. 
    Two tricritical wings due to the FM1 and FM2 states emerge from the tricritical point (TCP) under magnetic field and terminate at quantum wing critical points (QWCP). 
   The high-pressure magnetic (HP-MAG) state is represented by the yellow colored area.  
    It is noted that the HP-MAG state was initially considered as an antiferromagnetic state \cite{Taufour2016}, but the recent measurements suggest a short-range ferromagnetic ordered phase as shown in (b) \cite{Gati2021}.   
(b) Recent $T$-$p$ phase diagram in zero magnetic field for LaCrGe$_3$ after Ref. ~\cite{Gati2021}.
    The blue-shaded region corresponds to the region of FM order, which is schematically depicted in the insets by spins (arrows) pointing along the $c$ axis. 
    The thin and thick blue lines represent second and first order phase transition boundaries, respectively.  
     Dark and light yellow shaded regions relate to short-range FM ordered cluster phases that occur for $p$ $>$ 1.5 GPa. 
     The inset in this $p$  region visualizes  a possible short-range FM ordered phase.} 
\label{fig:Fig1}
\end{figure}

  In this context, LaCrGe$_3$ with  the hexagonal BaNiO$_3$-type structure [space group $P6_3/mmc (194)$]  \cite{Bie2007} is a peculiar itinerant ferromagnet because it was suggested to show both tricritical wings \cite{Kalu2017} and AFM state \cite{Taufour2016}  in the $p$-$T$-$H$ phase diagram as shown in Fig. \ref{fig:Fig1}(a).
   At ambient $p$, LaCrGe$_3$ is FM below the Curie temperature $T_{\rm C}$ $\sim$ 85~K  with an ordered magnetic moment  of 1.25 $\mu_{\rm B}$/Cr aligned along the $c$ axis at low temperatures \cite{Lin2013,Cadogan2013}.
   As shown in the phase diagram, upon reducing the temperature below $T_{\rm C}$, a crossover to a second FM state is observed around 70 K \cite{Kalu2017,Taufour2016}. 
   The two FM states, labeled FM1 and FM2 respectively, were also found in the itinerant ferromagnet UGe$_2$ \cite{Taufour2010} which were attributed to lower and higher values of saturated magnetic moment ($\mu_{\rm s}$)  respectively. 
   When $p$ is increased, $T_{\rm C}$ is suppressed and the second order FM transition becomes of the first order at $p\sim$ 1.8 GPa and $T$ around 40~K. 
 This yields a TCW  structure under $H$.  
   The FM1-FM2 crossover temperature is also suppressed under $p$ yielding a second TCW structure \cite{Kalu2017}.  
   It was also reported that, with increasing $p$,   the system exhibits  a new phase above 1.5 GPa \cite{Taufour2016}, which  was considered as a theoretically predicted AFM state.
   A second possible modulated AFM state was also suggested to appear  when further lowering temperatures at pressures higher than the critical pressure $p_{\rm c}\sim2.1$~GPa \cite{Taufour2016}. 
   These new phases are suppressed under $H$. 
  Quite recently, the detailed studies on LaCrGe$_3$ using thermodynamic, transport, x-ray, neutron scattering and $\mu$SR measurements reported  a modified  zero-field phase diagram [$T$-$p$ phase diagram, see Fig. \ref{fig:Fig1}(b)] \cite{Gati2021}. 
   Similar to the previous report \cite{Taufour2016},  a clear tricritical point has been reported with a slightly different position [$p$ = 1.5(1) GPa,  $T$ = 53(3) K]. 
   According to  Gati {\it et al.} \cite{Gati2021},  the phase appearing under high $p$ is likely not an AFM state but a form of disordered short-range FM clusters.   
   These results suggest that the avoided FM criticality in LaCrGe$_3$ has two features: (1) the change in the transition character from second order to first order  and (2) the appearance of short-range ferromagnetic  order rather than AFM order.
  Although the first one is a well-known mechanism for clean itinerant ferromagnets, the second one contradictorily suggests a sort of disorder in systems, making the system  peculiar.  
    Therefore, to understand the mechanism of the avoidance of FM criticality in LaCrGe$_3$, it is important to investigate the new phase  which is reported to appear under high pressures greater than $\sim$1.5 GPa. 
 
    With the motivation to investigate the evolution of magnetic properties of LaCrGe$_3$ under $p$ towards the putative QCP as well as to obtain more insights into the new phase under high pressures,  here we carried out nuclear magnetic resonance (NMR)  measurements  under $p$ up to  2.64 GPa.  
    Our previous $^{139}$La-NMR study  at ambient $p$ evidenced the presence of 3 dimensional (3D) isotropic ferromagnetic fluctuations in this system \cite{Rana2019}. 
   Furthermore, LaCrGe$_3$  was found to follow the generalized Rhodes-Wohlfarth (GRW) relation for 3D itinerant ferromagnets, and its location in the GRW plot suggested a high degree of localization in Cr 3$d$ electrons compared to other itinerant ferromagnets that exhibit the tricritical wings structure \cite{Rana2019, Lin2013}. 
   Our present NMR data show that the 3D FM fluctuations persist to dominate  in the paramagnetic state well above $T_{\rm C}$ in LaCrGe$_3$ throughout the measured $p$ region and  suggest a possible ferromagnetic order developing below $\sim$ 50 K under higher pressures in a magnetic field of $\sim$ 7.2 T.

 \section{Experimental Details}

   Rod-like shaped LaCrGe$_3$ single crystals were grown from high temperature solutions as detailed in Ref.~\cite{Lin2013}. 
The crystalline $c$ axis is parallel to the rod direction. 
   NMR measurements of $^{139}$La nuclei ($I$ = $7/2$, $\gamma_{\rm N}/2\pi$ = 6.0146 MHz/T, $Q=$ 0.21 barns) were carried out  using a lab-built phase-coherent spin-echo pulse spectrometer. 
Three large single crystals with a total mass of $\sim$150~ mg were aligned along the $c$ axis and were inserted with an NMR coil in a  NiCrAl/CuBe piston-cylinder for $p$ measurements up to 2.64 GPa. 
  Here the crystals were well separated by Teflon tapes and were placed in the NMR coil to make any demagnetization effects negligible. 
    Daphne 7474 was used as the $p$ transmitting medium and the $p$ calibration was carried out by measuring  the superconducting transition temperature of lead through resistivity measurements \cite{Eiling1981}. 
   The $^{139}$La NMR spectra were obtained by sweeping $H$ parallel to the $c$ axis at fixed resonant frequencies ($f$). 
    The zero-shift position corresponding to the Larmor field for each resonance frequency was determined by $^{31}$P NMR in H$_3$PO$_4$  solution or $^{63}$Cu NMR in Cu metal.    

  The $^{139}$La nuclear spin-lattice relaxation rate (1/$T_{\rm 1}$) was measured using a saturation recovery method.
   $1/T_1$ at each temperature  was determined by fitting the nuclear magnetization ($M$) versus time ($t$)  using the exponential function $1-M(t)/M(\infty) = 0.012e^{-t/T_1}+0.068e^{-6t/T_1}+0.206e^{-15t/T_1}+0.714e^{-28t/T_1}$,  where $M(t)$ and $M(\infty)$ are the nuclear magnetization at time $t$ after the saturation and the equilibrium nuclear magnetization at $t$ $\rightarrow$ $\infty$, respectively, for the case of magnetic relaxation \cite{Recovery}. 
    The observed recovery data in the paramagnetic state were well fitted by the function, indicating that the nuclear relaxation is mainly induced by fluctuations of the hyperfine field at the $^{139}$La site.

  \section{Results and discussion}

 \subsection{  $^{139}$La NMR spectrum} 
\begin{figure}[tb]
\includegraphics[width=\columnwidth]{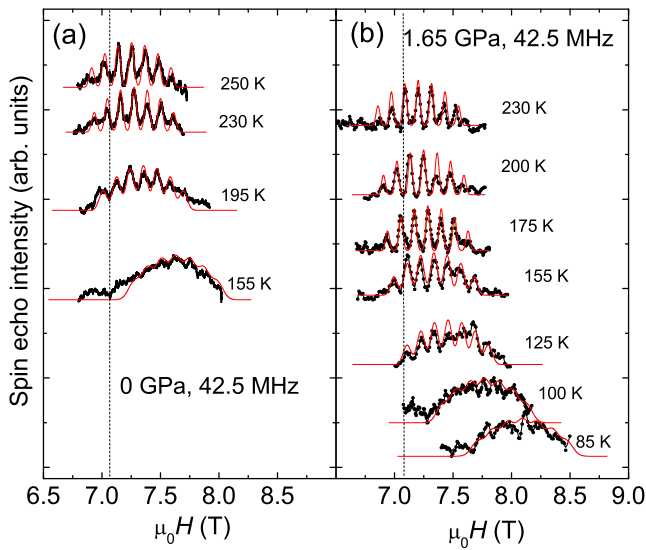} 
\caption{ Temperature dependence of field-swept $^{139}$La-NMR spectrum ($H$ $||$ $c$) under  (a) ambient pressure and (b) 1.65 GPa  measured at $f$ = 42.5 MHz. The black curves show  the observed spectra and the red curves are the calculated spectra with $\nu_{\rm Q}$ = 0.66 MHz and $\eta$ = 0 (see text). The vertical dashed black lines in (a) and (b) represent the zero-shift (Larmor) position ($\mu_{0}H_{0}=2\pi f/\gamma_{\rm N}$).}
\label{fig:Fig2}
\end{figure}

   $^{139}$La-NMR spectra in LaCrGe$_3$ at ambient pressure have been reported to show typical quadrupolar-split lines \cite{Rana2019} which are well explained by the combination of a large Zeeman interaction due to magnetic field  and a small quadrupole interaction whose nuclear spin Hamiltonian is given by 
${\cal{H}}=-\gamma\hbar\mathbf{I}\cdot\mathbf{H}_{\text{eff}}+\tfrac{h\nu_Q}{6}[3I_z^2-I^2+ \frac{1}{2}\eta(I_+^2 +I_-^2)]$.
Here  $\mathbf{H}_{\text{eff}}$ is the effective magnetic field at the La site, $h$ is Planck's constant,  $\nu_{\rm Q}$ is nuclear quadrupole frequency defined by $\nu_{\rm Q} = 3e^2QV_{ZZ}/2I(2I-1)h$  where $Q$ is the electric quadrupole moment of the La nucleus, $V_{ZZ}$ is the electric field gradient (EFG) at the La site, and $\eta$ is the asymmetry parameter of  EFG at the La site.
   From the angle dependence of $^{139}$La NMR spectra in the paramagnetic state at ambient pressure, the principal axis of the EFG was determined to be parallel to the $c$ axis and $\nu_{\rm Q}$ is estimated to be  0.66 MHz with $\eta$ $\sim$ 0 \cite{Rana2019}. 
    In fact  the observed spectra are well reproduced by the calculated spectra [red curves in Fig. \ref{fig:Fig2}(a)] from the Hamiltonian with those parameters \cite{Rana2019}.     
   As in the case of the ambient pressure, $^{139}$La-NMR spectra in LaCrGe$_3$ under pressure show the typical $I$ = 7/2 quadrupole split lines with a nearly pressure independent value of $\nu_{\rm Q}$ = 0.66 MHz and  $\eta$ $\sim$  0 in the paramagnetic state well above the magnetic ordering temperature under pressure up to the highest measured $p$ of 2.64~GPa.

\begin{figure}[tb]
\includegraphics[width=\columnwidth]{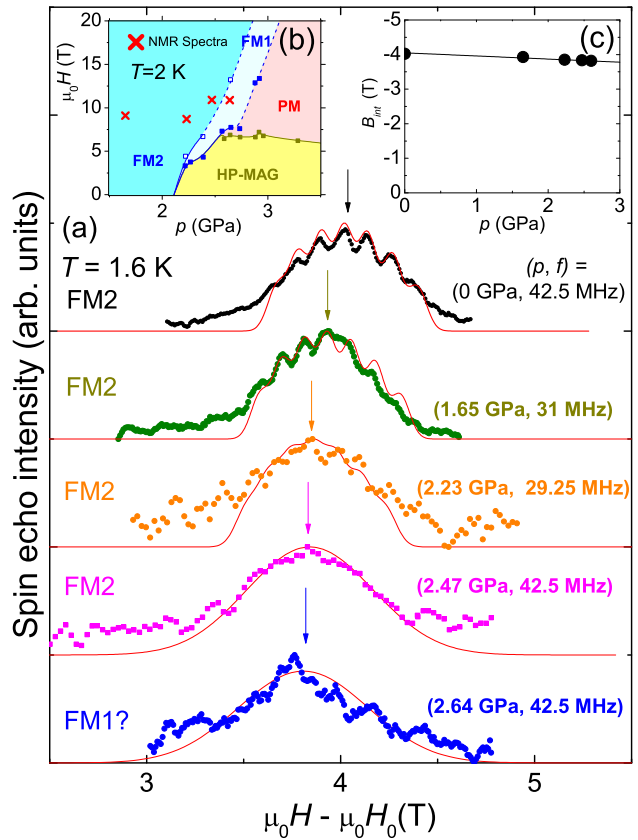} 
\caption{(a) Field-swept $^{139}$La-NMR spectra of LaCrGe$_3$ at 1.6 K for various pressures up to 2.64 GPa, as a function of the difference in external    magnetic field ($H$) and resonance field ($f/\gamma$). 
   The red curves are the calculated spectra with appropriate magnetic broadening.  
   The downward arrows represent the peak position of the central peak for each pressure determined from the calculated spectrum. 
  Inset: (b) $H$-$p$ phase diagram at $T=2$~K taken from Ref. \cite{Kalu2017}. 
    The red crosses represent the positions of the central peak of the NMR spectra.  
(c) Internal magnetic induction ($B_{\rm int}$)  at the La site   as a function of $p$. The black line represents a linear fit.}
\label{fig:Fig3}
\end{figure}

   As shown in the typical temperature dependence of the field-swept $^{139}$La NMR spectrum at $p$ = 1.65 GPa for $H$ parallel to the $c$ axis ($H$ $||$ $c$) [Fig. \ref{fig:Fig2}(b)],  with decreasing temperature, each line becomes broader due to inhomogeneous magnetic broadening and the spectra show less clear features of the quadrupolar split lines below $T$ $\sim$ 100~K.
   A similar broadening of NMR lines was reported for the case of ambient pressure \cite{Rana2019} as shown in Fig. \ref{fig:Fig2}(a). 
   It is noted that one can see the well-split lines even at 155 K at $p$ = 1.65 GPa although the featureless spectrum was observed at the same temperature  at ambient pressure. 
   This indicates that $T_{\rm C}$ decreases with the application of pressure, consistent with the reduction in $T_{\rm C}$ from 85 K at ambient pressure to $\sim$50 K at $p$ = 1.65 GPa determined by the resistivity measurements \cite{Taufour2016}.
    The broadening of the spectra at a wide range of temperatures close to $T_{\rm C}$  make the NMR spectrum measurements difficult around $T_{\rm C}$.
     However, when the temperature is decreased down to 1.6 K, well below $T_{\rm C}$, we were able to observe the $^{139}$La NMR spectrum in the FM state.
   Figure\ \ref{fig:Fig3} shows the field-swept $^{139}$La-NMR spectra of LaCrGe$_3$ at $T$ = 1.6 K ($H||c$) at various pressures from $p$ = 0 GPa up to 2.64~GPa, where all the spectra are largely shifted from the Larmor field (2$\pi f$/$\gamma_{\rm N}$ = $\mu_{\rm 0}H_{0}$) to a  higher magnetic field by $\sim$ 4 T. 
  The shift is due to the internal magnetic induction ($B_{\rm int}$$\sim$ -4 T) at the La site produced by the Cr ordered moments in the FM state, as reported from the NMR measurements at ambient pressure in Ref. \cite{Rana2019}.  
      Note that the horizontal axis of each spectrum is shifted by $\mu_{\rm 0}H_{0}$ and now corresponds to the negative value of $B_{\rm int}$. 
    As shown in Fig. \ref{fig:Fig3}, although the clear feature of quadrupole split lines of the spectrum can be seen at ambient pressure, the peak structures become less prominent with increasing $p$ where the line width [determined by the full width at half maximum (FWHM)] increases from  $\sim$0.67 T at $p$ = 0 to $\sim$ 0.9 T at the high pressure region ($P$ $>$ 2.23 GPa).           
     This suggests that a sort of inhomogeneity is induced in the system by the application of pressure. 
    It would be interesting if the inhomogeneity observed in the NMR measurements corresponds to the possible disorder in LaCrGe$_3$ under pressure inferred from the short-range magnetic ordered state \cite{Gati2021}.  
    As can be seen in the figure, the peak positions of the spectra, marked by downward arrows, shift slightly to lower magnetic fields due to the decrease in $|B_{\rm int}|$. 
     The pressure dependence of $B_{\rm int}$ is shown in Fig. \ref{fig:Fig3}(c) as a function of $p$.
    The $|B_{\rm int}|$ decreases slightly by only $\sim$ $5\%$ up to 2.64 GPa. 
     These results seem to be consistent with the results of $\mu$SR measurements where $B_{\rm int}$ at the muon site in the FM state is nearly independent of pressure \cite{Taufour2016}.
  According to the $H-p$ phase diagram at 2 K reported in Ref.~\cite{Kalu2017} shown in Fig. \ref{fig:Fig3}(b),  our $^{139}$La NMR spectrum at $p$ = 2.64 GPa would correspond to the one in the FM1 state, whereas other spectra with different pressures represent the ones in the FM2 state.
  As seen in Fig. \ref{fig:Fig3}, we observed no significant change in $B_{\rm int}$ between  2.47 GPa and 2.64 GPa. 
  Since the internal field $B_{\rm int}$  is proportional to the spontaneous magnetization, the results suggest that either $\mu_{\rm s}$ does not show prominent change between FM1 and FM2, or that we did not cross the FM1-FM2 boundary in our experiment, as will be further discussed later.


\begin{figure}[tb]
\includegraphics[width=\columnwidth]{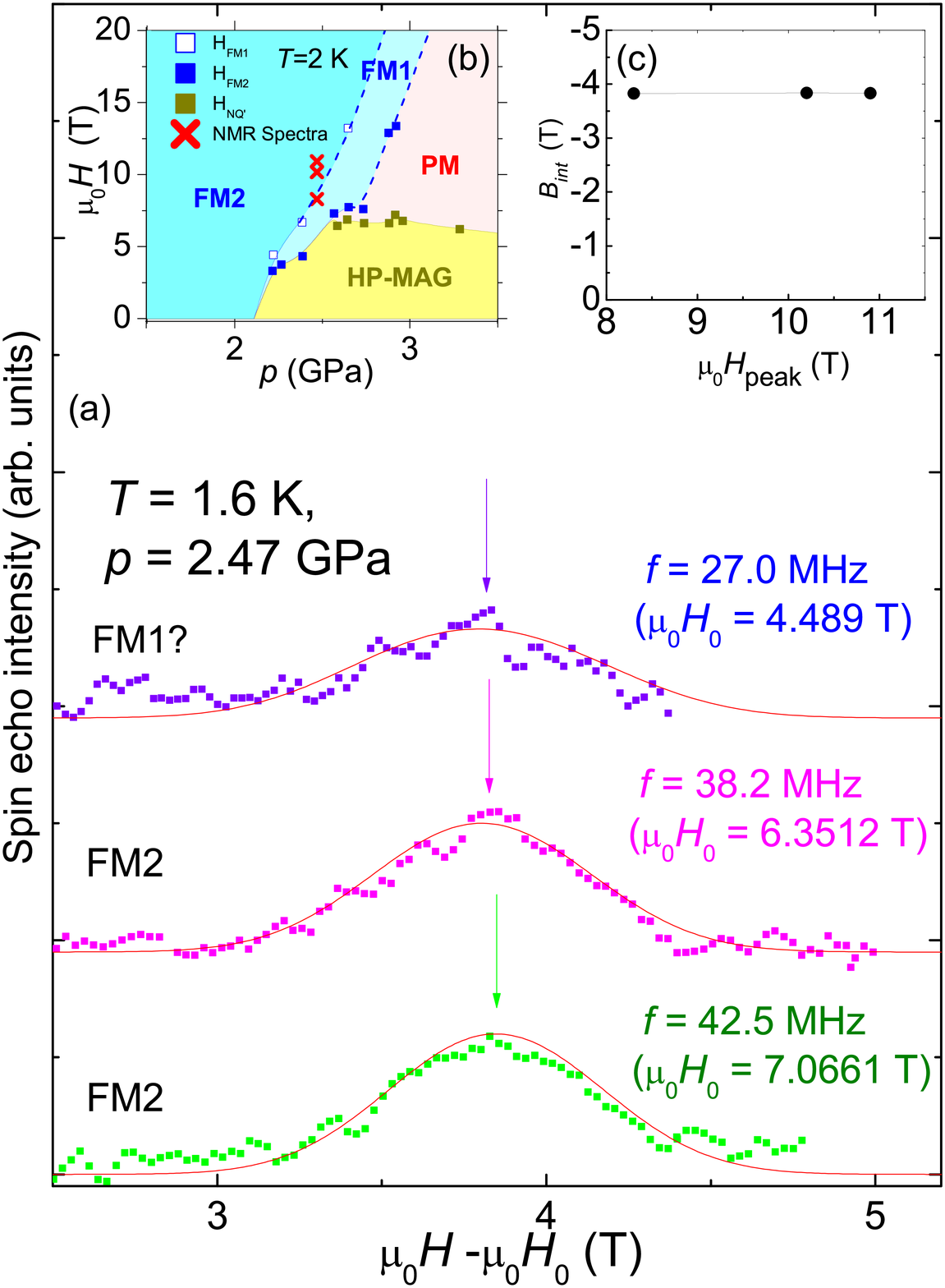} 
\caption{(a) $^{139}$La NMR spectra at $p$ = 2.47 GPa taken at various frequencies, as a function of the difference in external field ($\mu_{0}H$) and the zero shift position ($\mu_{0}H_{0}=2\pi f/\gamma_{\rm N}$) for $H||c$ direction at $T$ = 1.6 K.  
 The downward arrows show the peak positions of the central transition lines of the spectra and the red lines which serve as guides for the eye have the same full width at half maximum (FWHM).
 Inset: (b) $H$-$p$ phase diagram at $T=2$~K taken from Ref. \cite{Kalu2017} with the positions (red crosses) where the central peak of the NMR  
spectra was measured. (c)  $H_{\rm peak}$ dependence of $B_{\rm int}$ for  $H||c$ for the three different resonance frequencies. 
}
\label{fig:Fig4}
\end{figure}

   Since $B_{\rm int}$ also depends on a hyperfine coupling constant ($A$) which may change by the application of pressure, one could have a chance to accidentally observe no change in $B_{\rm int}$ due to a compensation of the changes in both $\mu_{\rm s}$ and $A$, although the large change in $A$ is unlikely
 since we do not see any significant change at high temperatures above $\sim$200 K at various $p$ in nuclear spin-lattice relaxation times ($T_1)$ which are proportional to the square of $A$, as will be described below.
   Therefore we have measured the magnetic field dependence of NMR spectra at a constant pressure of 2.47~GPa by changing the resonance frequencies from 27 MHz to 42.5 MHz. 
    In this case, any change in $A$ would not be expected. 
    The observed spectra are shown in Fig. \ref{fig:Fig4} where the peak positions of the spectra  increase from $\mu_0 H_{\rm peak}$ = 8.3~T at the lowest resonance frequency of $f$ = 27~MHz up till $\mu_0 H_{\rm peak}$ = 10.9~T at $f$ = 42.5~MHz.       
    Note that the horizontal axis of each spectrum is shifted  by the Larmor field as in the case of Fig. \ref{fig:Fig3}.
    From the reported $H-p$ phase diagram at $T$ = 2 K \cite{Kalu2017} [(Fig. \ref{fig:Fig4}(b)] where one  expects the FM1-FM2 crossover around 9.2 T, the NMR spectrum at 27.0~MHz may be considered to represent the FM1 state whereas those measured at the higher frequencies represent the FM2 state.
   It is found that $B_{\rm int}$ $\sim$ $-$3.82 T is nearly independent of the external fields as shown in Fig. \ref{fig:Fig4}(c) 
even though $H$ crosses the border between the FM1 and FM2 states at 9.2 T in Ref. \cite{Kalu2017}. 
 We also do not observe any clear change in the line width which reflects the distribution of both the magnitude and direction of $\mu_{\rm s}$.
  While the signal intensity decreases with decreasing the resonance frequency, the line width at 27.0 MHz is similar to that at 38.2 MHz within our experimental uncertainty, as shown by the red curves describing the broad peak with the same line width. 
   Thus, either no clear change in the magnetic states between FM1 and FM2 is detected, or we did not cross the FM1-FM2 boundary in our experiments, as will be discussed below.



   The above discussions are based on  the phase diagram reported in Refs. \cite{Kalu2017,Taufour2016, Taufour2018} [see, Fig. \ref{fig:Fig1}(a)], however, 
it is important to point out that the FM1-FM2 crossover field could vary depending on different experimental set up as well as sample quality.
  The $H$-$p$ phase diagram shown in Figs. \ref{fig:Fig3}(b) and \ref{fig:Fig4}(b) was determined by the  measurements with modified  Bridgman pressure cells which may produce larger pressure gradients compared with piston-cylinder pressure cells. 
  The ferromagnetic instability in LaCrGe$_3$ has been shown to be primarily driven by the Cr-Cr distance along the $c$ axis \cite{Nguyen2018}, which would be the direction of higher pressure gradients in the previous measurements in Bridgman cells \cite{Kalu2017,Taufour2018}.
  Therefore, the results of the NMR measurements by utilizing the piston-cylinder pressure cells should be  compared with the phase diagram of Fig. \ref{fig:Fig1}(a) with some caution. 
    It is possible that our NMR experiments did not reach the FM1-FM2 crossover field. 
   To discuss the magnetic properties of the FM1 and FM2 phases, it is important to make sure that the system is actually in each phase.     
   Since the FM1-FM2 crossover in the compound has been detected by resistivity measurements \cite{Taufour2016}, one should measure the resistivity of the same sample in the same experimental set up (i.e., pressure cells). 
   This would require performing simultaneous measurements of NMR and insitu resistivity. 
This is beyond the scope of the present work and is planned for a future work.

 \subsection{$^{139}$La-NMR shifts}

\begin{figure}[tb]
\includegraphics[width=7.5cm]{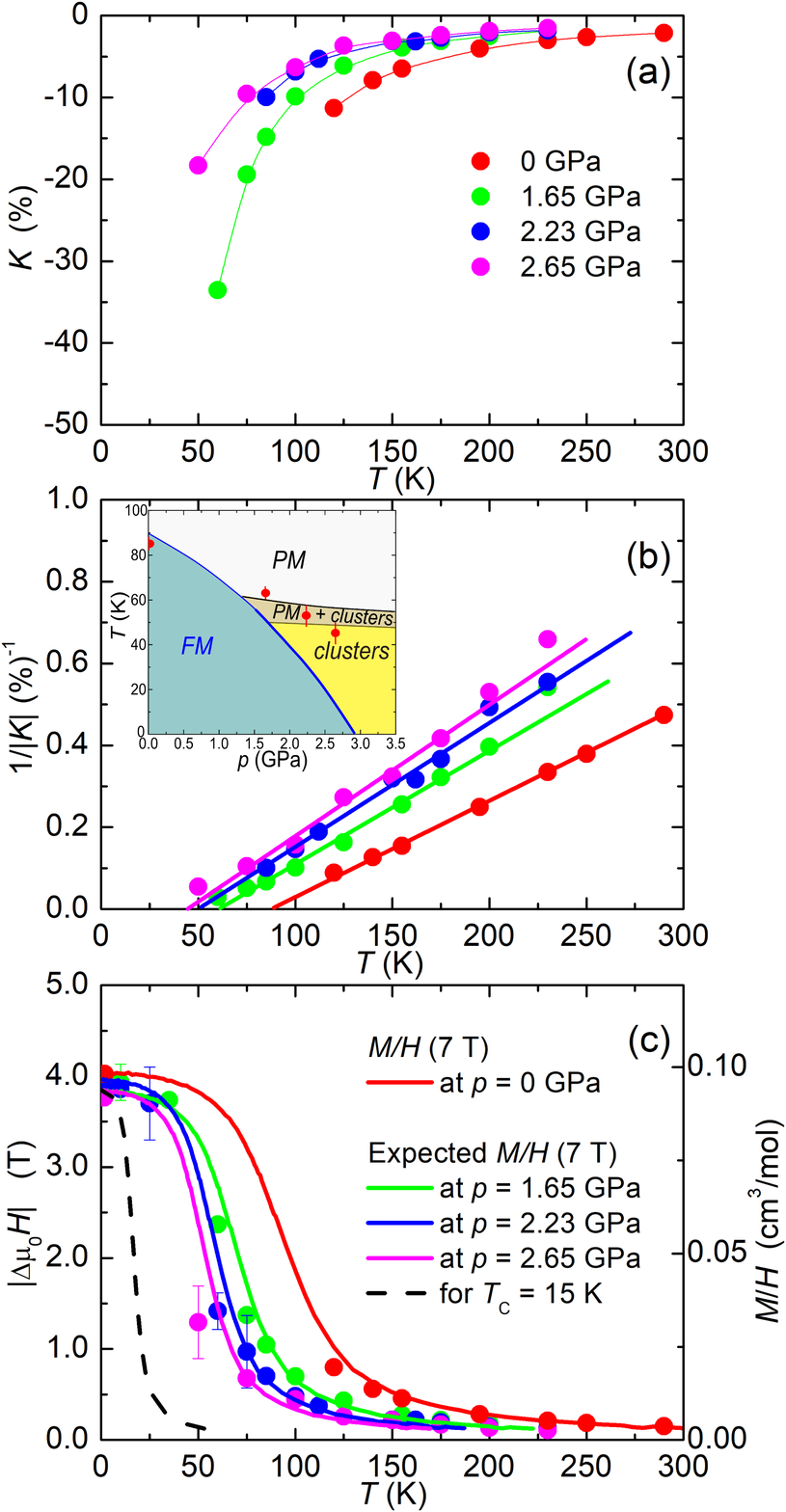} 
    \caption{(a) Temperature dependence of  $^{139}$La Knight shift $K$ for various pressures in the $H||c$ direction. 
   (b)     The temperature dependence of 1/$|K|$ for various $p$ values. 
   The inset shows the zero-field phase diagram [Fig. 1(b)] together with the estimated $T_{\rm C}$ at $H$ $\sim$7 T shown by the red  circles.
   (c) Temperature dependence of $\Delta \mu_0 H$ for various pressures. The red curve shows the temperature dependence of $M/H$ measured at 7 T at ambient pressure.  Other curves (solid lines) show the $M/H$ data   estimated from the ambient pressure 7 T data shown (see text for details).
The black dashed curve shows the expected temperature dependence of $M/H$ with $T_{\rm C}$ = 15 K under a magnetic field of 7 T. }
\label{fig:Fig5}
\end{figure}

   The $T$ dependence of $^{139}$La NMR Knight shift ($K$) is shown in Fig.~\ref{fig:Fig5}(a)  for the measured pressures with the $H||c$ direction. 
    The $K$s show the Curie-Weiss (CW) type temperature dependence and $K$ decreases monotonically with decreasing temperature.    
    With increasing $p$, the CW type behavior of $K$ shifts to lower temperatures, again consistent with the suppression of $T_{\rm C}$.  
 
    In general, $K$ has contributions from the temperature  dependent spin shift $K_{\rm s}(T)$ and $T$ independent chemical shift $K_{\rm 0}$: $K(T)$ = $K_{\rm s}(T)$ + $K_{\rm 0}$ where $K_{\rm s}(T)$ is proportional to the spin part of magnetic susceptibility  $\chi_{\rm s}$($T$) via hyperfine coupling constant ($A$), $K_{\rm s}(T)$  =  $A\chi_{\rm s}(T)/N_{\rm A}$ where $N_{\rm A}$ is Avogadro's number. 
    At ambient pressure, from the $K$ vs. $\chi$ plot analysis,  the $A$ was estimated  to be $-27$~kOe/$\mu_{\rm B}$ for $H||c$ \cite{Rana2019}. 
    In addition,  $K_{\rm 0}$ was reported to be close to zero \cite{Rana2019}, indicating that the observed $K$ is mainly attributed to $K_{\rm s}(T)$.
      
    Since the $\chi_{\rm s}$ at ambient pressure has been reported to follow the CW-type behavior even though the system is itinerant \cite{Bie2007, Lin2013,Rana2019,Taufour2018}, one may estimate $T_{\rm C}$  from the temperature dependence of $K$ since $K$ at high temperatures  is proportional to $\chi_{\rm s}$ which is proportional to  $C$/($T-T_{\rm C}$) ($C$: Curie constant).
    Fig. \ref{fig:Fig5}(b) shows the temperature dependence of 1/$|K|$ where the intercepts of the $x$ axis provide an estimate of $T_{\rm C}$ for each pressure. 
    The intercept for the case of ambient pressure is estimated to be $\sim$85 $\pm$ 3  K which is in excellent agreement with $T_{\rm C}$ = 85 K reported previously.
   From the intercepts, we estimated the $p$ dependence of $T_{\rm C}$ which decreases from 85 K at ambient pressure to 63 $\pm$3 K at 1.65 GPa, to 53 $\pm$ 5 K at 2.23 GPa, and to  45 $\pm$ 5 K at 2.64 GPa.
   It is noted that the slopes in the 1/$K$ vs. $T$ plots slightly increase with increasing $p$.  
   Assuming the change were only due to the change in Curie constant (that is, no change in $A$), the effective moments are estimated to slightly decrease by at most 14 \% at $p$ = 2.64 GPa from the value at ambient pressure.    
    It is also noted that no clear signature of antiferromagnetic interaction can be found from the temperature dependence of $K$ where all intercepts are positive for the measured $p$ region.

   We also checked the estimated $T_{\rm C}$ with  the temperature dependence of $B_{\rm int}$ as well as the temperature dependence of $H$-induced hyperfine field at the La site  by using magnetization ($M/H$) data.
   The red circles in Fig. \ref{fig:Fig5}(c) show the temperature dependence of $|$$\Delta \mu_0 $$H$$|$ = $\mu_0$$H_{\rm peak}$ -$\mu_0$$H_0$  (corresponding to $B_{\rm int}$ in the FM state and to the $H$ induced hyperfine field at the La site in the PM state) at ambient pressure which is expected to be proportional to the magnetization $M$ in both the PM and FM states in ferromagnets. 
   In fact, the temperature dependence of  $|$$\Delta \mu_0 $$H$$|$  measured at ambient pressure seems to be well reproduced by the $M/H$ curve (red curve) measured at  $\mu_0$$H$ = 7.0 T, although no data points are available in a wide temperature range around $T_{\rm C}$.  
   The temperature dependence of  $|$$\Delta \mu_0 $$H$$|$  for other pressures shown in the figure was also reasonably reproduced for not only the high temperature region but also the low temperature region by the corresponding solid   $M/H$ curves.  
   Since no $M/H$ data under 7 T are available for pressure other than ambient  at present, which will be requested to measure in the future, we estimated the $M/H$  behavior for other pressures based on the ambient data under 7 T. 
   The solid curves in Fig. \ref{fig:Fig5}(c)  are the expected $M/H$ curves  where the temperature for each pressure is normalized to the estimated $T_{\rm C}$ [i.e.  $T$($T_{\rm C}$/$T_{\rm C}$($p$=0))] and the magnetization is also scaled to the lowest temperature $\Delta \mu_0 $$H$ data point.
   Thus we conclude that the estimation of $T_{\rm C}$ from NMR measurements seems to be reasonable.

 According to the phase diagram shown in Fig. \ref{fig:Fig1}(b), the long-range FM state appears under $H$= 0  at $T$ $\sim$  30~K and 15~K for $p$ = 2.26 and 2.64 GPa, respectively.
   Thus it is important to check whether or not  the observed temperature dependence of $M$ ($\propto$ $|$$\Delta$$H$$|$) can be explained by the $T_{\rm C}$   under magnetic field, as the application of magnetic field produces a large tail of magnetization even above $T_{\rm C}$. 
    As shown in Fig. \ref{fig:Fig5}(c), the observed temperature dependence of $|$$ \Delta \mu_0$$H$$|$ at $p$ =  2.64 GPa cannot be explained at all by the black dotted curve which is the expected $M/H$ behavior under 7 T for the reported zero-field $T_{\rm C}$ = 15 K at this pressure \cite{Gati2021}.     
  Therefore,  the higher $T_{\rm C}$ obtained by the NMR measurements under high pressures cannot be explained by an artificial effect of the application of magnetic field, and those values may suggest the increase of $T_{\rm C}$ by the application  of magnetic field.
   It is also interesting to note that the estimated $T_{\rm C}$s from the NMR measurements are close to the transition temperatures for the short-range FM order phase or a mixed state of the FM cluster and PM states determined at $H$ = 0 [see  the inset of Fig. \ref{fig:Fig5}(b)]. 
   Therefore, another possibility is that the NMR spectrum measurements detect the short-range FM ordered state. 
    Since we apply magnetic field for the NMR measurements, a ferromanetic phase transition is smeared, therefore, $T_{\rm C}$ cannot be well defined and  only the crossover temperature from a paramagnetic state to a ferromagnetic state in magnetic  field can be inferred.
   This  makes it difficult to distinguish between long-range and short-range ordered states by measuring NMR spectra.  
Nevertheless, we may conclude that our NMR data suggest that FM orderings start to develop below $T \sim 50$ K under $p$ $>$ $\sim$1.5 GPa and $\mu_0$$H$ $\sim$ 7.2 T.


 \subsection{Magnetic fluctuations} 



\begin{figure}[tb]
\includegraphics[width=\columnwidth]{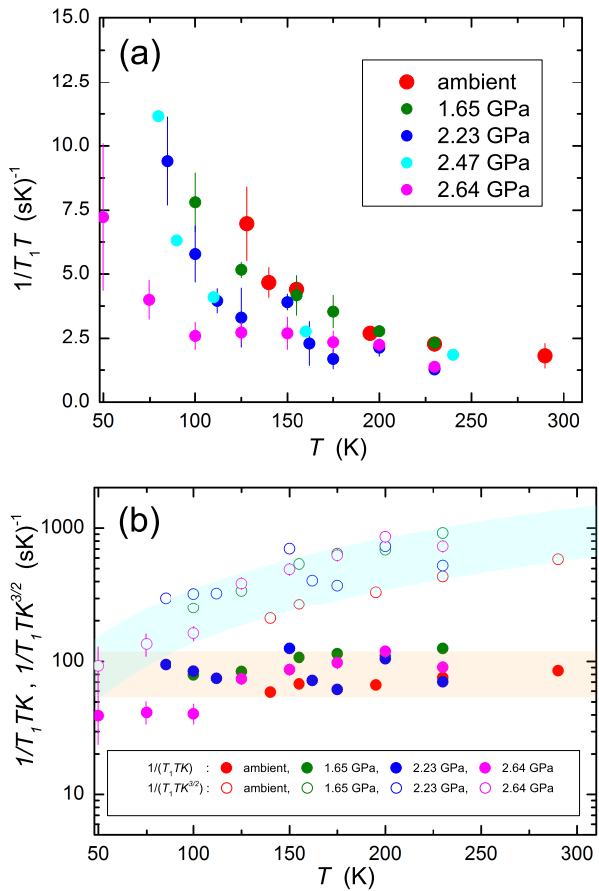} 
\caption{(a) Temperature dependence of $^{139}$La 1/$T_1T$ under various pressures in the PM state measured at $\mu_0$$H$ $\sim$ 7.2 T. 
(b) Temperature dependence of  1/($T_1TK$)  and 1/($T_1TK^{3/2}$) in the PM state measured at $\mu_0$$H$ $\sim$ 7.2 T. 
 }
\label{fig:T1}
\end{figure}

In order to investigate the magnetic fluctuations in  LaCrGe$_3$ and their evolution with $p$, we measured the temperature dependence of $^{139}$La spin-lattice relaxation rate (1/$T_1$) at the peak position of the spectra at various pressures with the $H||c$ direction. 
  Figure \ref{fig:T1}(a) shows 1/$T_1T$ as a function of $T$ for various $p$ values. 
  At all pressures, 1/$T_1T$ increases with lowering temperature from room temperature down towards $T_{\rm C}$. 


    In our previous paper \cite{Rana2019}, we discussed FM magnetic fluctuations in LaCrGe$_3$ based on the $T_1$ and $K$  data at ambient pressure using  the self-consistent renormalization (SCR) theory. 
    Here we analyze the present NMR data under pressure with the same theory.
     According to the SCR theory for weak itinerant ferromagnets,  1/($T_1TK_{\rm s}$) and 1/($T_1TK^{3/2}_{\rm s}$) are expected to be independent of $T$  for three dimensional (3D) or  two-dimensional (2D) FM spin fluctuations, respectively \cite{Moriya1974,Hatatani1995}. 
   Utilizing the $1/T_1T$ data in the PM state well above $T_{\rm C}$ for various pressures shown in Fig. ~\ref{fig:T1}(a), we plotted the $T$ dependence of 1/($T_1TK$) and 1/($T_1TK^{3/2})$ for various pressures in Fig.~\ref{fig:T1}(b).
      As in the case of ambient pressure, $1/T_1TK$ for each pressure seems to be nearly constant in the high temperature region, while $1/(T_1TK^{3/2})$ decreases slightly with decreasing $T$ at all $p$ values. 
    Therefore, we conclude that the magnetic fluctuations in this system are dominated by 3D FM fluctuations through the measured pressures up to 2.64 GPa, indicative of the robust nature of ferromagnetism in LaCrGe$_3$.   
    It is noted that the $1/T_1TK$ data below 100 K for $p$ = 2.64 GPa seem to deviate from the constant value at temperatures higher 125 K.
Although the reason for the deviation is not clear at present, the decrease in the value of  the 1/$T_1TK$ below 125 K suggests the suppression of 3D FM spin fluctuations. 
   It is also noted that, in the recent reported phase diagram under zero magnetic field \cite{Gati2021},  $p$ = 2.64~GPa is close to the pressure where the first order FM transition is completely suppressed.   
   Further studies  are required to see how magnetic fluctuations change with $p$  beyond 2.64~GPa.


\section{Summary}
   In conclusion, we carried out $^{139}$La nuclear magnetic resonance measurements in the itinerant ferromagnet LaCrGe$_3$ under pressure to investigate its static and dynamic magnetic properties. 
   From the pressure dependence of $^{139}$La-NMR spectra at $T$ =1.6 K, the internal magnetic induction  $B_{\rm int}$ at the La site in the ferromagnetic ordered state is found to decrease very slightly  by less than 5\%  with $p$ from ambient pressure to 2.64 GPa. 
  This indicates that  the Cr 3$d$ ordered moments are robust under pressure.
   In the ferromagnetic state, we observed the broadening of NMR spectra under high pressures above  $p$ $>$ 2.23 GPa. 
   This suggests that inhomogeneity is induced by the application of pressure, which could be consistent with a possible disorder in this system under pressure as pointed out in Ref. \cite{Gati2021}.
   In addition, from the temperature dependence of $B_{\rm int}$ and the Knight shift, the ferromagnetic state is revealed to exist below $\sim$ 50 K at $p$ = 2.23 and 2.64 GPa under a magnetic field of $\sim$ 7.2 T although we could not distinguish between long-range or short-range magnetic order states.  
   Based on the analysis of NMR data using the self-consistent-renormalization theory, the spin fluctuations in the paramagnetic state well above $T_{\rm C}$ are revealed to be three dimensional ferromagnetic throughout the measured $p$ region. 
  In this sense, as pointed out in Ref. \cite{Gati2021}, LaCrGe$_3$ might  stand as a peculiar system having a new route to avoid a ferromagnetic quantum critical point by not only changing the order of the phase transition but also through the appearance of  the high-pressure magnetic phase probably dominated by ferromagnetic interactions. 
   To understand the nature of the avoidance of ferromagnetic quantum criticality in  LaCrGe$_3$, further detailed studies  under lower magnetic fields as well as higher pressures greater than $\sim$3~GPa  will be required.

\section{Acknowledgments}

   The authors would like to thank Y. Kuwata and Y. Noma  for their help in conducting experiments and Q.-P. Ding for valuable discussions.      
  The research was supported by the U.S. Department of Energy, Office of Basic Energy Sciences, Division of Materials Sciences and Engineering. 
    Ames Laboratory is operated for the U.S. Department of Energy by Iowa State University under Contract No.~DE-AC02-07CH11358.
      Part of the work was supported by the Japan Society for the Promotion of Science KAKENHI Grant Numbers JP15H05882, JP15H05885, JP15K21732, and  JP18H04321 (J-Physics).
     K. R. also thanks the KAKENHI: J-Physics for the financial support that provided an opportunity to be a visiting scholar at Kobe University.

\end{document}